\documentstyle[12pt,epsbox]{article}
\begin{document}
\baselineskip 5mm

\title{QCD Monopoles and Chiral Symmetry Breaking
on SU(2) Lattice}

\author{O. Miyamura and S. Origuchi \cr
Department of Physics, Hiroshima University,\cr
Higashi-Hiroshima 724, Japan
\thanks{E-mail address: miyamura@fusion.sci.hiroshima-u.ac.jp}}
\date{}
\maketitle

\begin{abstract}
Pseudoscalar correlator is measured in a singular (monopole dominant) and
a regular (photon dominant) parts of a maximal abelian field on SU(2)
lattice. In the abelian field and its singular part, light pseudoscalar
boson are observed similar to that in SU(2) field.
On the other hand , the correlator in the regular part behaves
like a product of free quark and anti-quark.
Obtained results give a support for a possibility
that monopole condensation is responsible for chiral symmetry breaking
as well as confinement.  Correlation between monopoles and
instantons is also investigated by observing topological charge.
A field including monopoles carrys topological charge with good correlation
to that in original SU(2) gauge field
while that including only photons shows no topological charge.
\end{abstract}

\section{Monopole condensation and chiral symmetry breaking}
Recently,
chiral symmetry breaking
in the maximal abelian field and in its monopole-part
has been examined
on SU(2) lattice at finite temperature by one of the present authors.
\cite{Miya}
It has been found that
chiral condensates (in quench level) in the U$_1$ field
and in its monopole part ,
keep the same feature of chiral breaking
with that observed in original SU(2) gauge field.
Existence of light pion in the maximal abelian field has also
been reported at zero temperature.\cite{Wolo}

As for confinement, condensation of monopoles has got
increasing evidences.
Wriggling of magnetic current in confinement phase\cite{Kron}
abelian dominance in confinement,\cite{Brands,Hioki}
saturation of area law of Wilson loops by monopole
contribution,\cite{Ohno}
evidence of monopole condensation by
entropy dominance over energy of magnetic current \cite{Shiba}
and finiteness of an order parameter in confinement phase\cite{Giacomo}
have been reported.

Those observations give a support for an idea that
monopole condensation is responsible for chiral symmetry breaking
as well as confinement.
Historically, such idea has been
discussed several times.  Banks and Casher first arged that
confinement breaks chiral symmetry in 1980.\cite{Casher}
Later, some authors tried to show chiral symmetry breaking
based on dual Ginzburg-Landau theory.\cite{Baker,Klein,Kamizawa}
A recent work by Suganuma, Sasaki and Toki
has given an affirmative answer for this possibility.\cite{Suga}

Here we investigate pseudoscalar boson in the field dominated
by monopoles on lattice for a further investigation.
We study relationship between instantons and monopoles
also by observing topological charge.
Present analyses have been done
using Intel Paragon
XP/S(56nodes) at the Institute for
Numerical Simulations and Applied Mathematics in Hiroshima University.

\section{ U$_1$ gauge field by maximal abelian projection and
its decomposition into monopole and photon parts}

{}From an SU(2) gauge field, we extract U$_1$ field through maximal abelian
projection\cite{tHooft2}
by maximizing

$$
R= \sum _{n,\mu} Tr(\sigma_3U_{\mu}(n)
\sigma_3U_{\mu}(n)^{\dagger}) , \eqno[1]
$$

and spliting SU(2) gauge link $U_{\mu}(n)$ as

$$
U_{\mu}(n) = \left[\matrix{r_c(n)  & -c_{\mu}(n)
\cr c_{\mu}(n)^* &
r_c(n) }\right ] \times
\left[ \matrix{ e^{i\theta^{}_{\mu}(n)} & 0 \cr  0 &  e^{-i\theta^{}_{\mu}(n)}
} \right ] , \eqno[2]
$$

where $r_c(n)=\sqrt{1-|c_{\mu}(n)|^2}$.
Here, $\theta^{}_{\mu}(n)$ is the U$_1$ gauge field and
$c_{\mu}(n)$ is interpreted as a charged matter field under the U$_1$
gauge transformation.

Decomposition into the singular(monopole) and the regular(photon)
parts is carried out following Matsubara et al.
\cite{Matsu1,Suzuki}
On lattice , every quantity is regular at finite lattice spacing and we
can work in Landau gauge. Then we have Landau gauge field
as

$$
     \theta^L_{\mu}(n)=\sum _m G(n-m) \partial_{\lambda}
\theta^{}_{\lambda\mu}(m) , \eqno[3]
$$

where the abelian field strength $\theta^{}_{\mu\nu}$ is given by
$
\theta^{}_{\mu\nu} =
\partial_{\mu}\theta^{}_{\nu}-\partial_{\nu}\theta^{}_{\mu}$
and $G(n)$ is lattice Coulomb propagator and $\partial$ is
derivative on lattice.

Here we decompose field strength following
DeGrand-Toussaint as,\cite{DeG}

$$
  \theta^{}_{\mu\nu}=\bar \theta^{}_{\mu\nu} + 2\pi M_{\mu\nu}
, \eqno[4]
$$

where $ -\pi < \bar \theta^{}_{\mu\nu} < \pi$ and
$M_{\mu\nu}$ is Dirac string.

Substituting eq.[4] into eq.[3], we have a decomposition of
$\theta^L_{\mu}(n)$
into a regular part $\theta_{\mu}^{Ph}(n)$ and a singular
part $\theta_{\mu}^{Ds}(n)$
where

$$
     \theta_{\mu}^{Ph}(n) = \sum _m G(n-m) \partial_{\lambda}
\bar \theta^{}_{\lambda\mu}(m) , \eqno[5]
$$

and

$$
     \theta_{\mu}^{Ds}(n) = 2\pi \sum _m G(n-m) \partial_{\lambda}
M_{\lambda\mu}(m) . \eqno[6]
$$

It is noted that $\theta_{\mu}^L(n)=\theta_{\mu}^{Ph}(n)+\theta_{\mu}^{Ds}(n)$.
Such decomposition has been discussed
by de Forcrand et al.\cite{Forc}

As a result of the decomposition, the singular part carrys almost
equal number of magnetic currents ($95 - 98\%$) to
that in the $U_1$ field whereas
the number of electric currents is less than
a few $\%$ of that in the $U_1$
field. On the other hand, situation is
just opposite in the regular part.\cite{Miya}
By this reason,
the singular part is called as monopole part and
the regular part as photon part.

\section{ Pseudoscalar correlator in the field dominated by monopoles}

Existence of light pseudoscalar boson is examined to confirm chiral broken
phase in U$_1$ and the singular gauge field $\theta_\mu (n)^{Ds}$.
We measure pseudo
scalar correlator by staggered quark on $16^3\times 32$ lattice
at $\beta=2.2$.

The quark operator is given by

$$
D=-ma\delta_{n,m}-\sum_{\mu}\eta (n)[ V_{\mu}(n)\delta_{n+\mu,m}-
V_{\mu}^{\dagger}(n-\mu)\delta_{n+\mu,m}], \eqno[7]
$$

and
$V_{\mu}(n)$ is set to be either $U_{\mu}(n)$ or
$e^{i\theta_{\mu}(n)}$ or
$e^{i\theta_{\mu}^{Ds}(n)}$ or
$e^{i\theta_{\mu}^{Ph}(n)}$.
Quark mass parameter is taken as $ma=0.005$
and number of configurations is 35.

Results are presented in Fig.1.
The pion correlators in SU(2), U$_1$ and
$\theta_\mu (n)^{Ds}$ gauge fields have similar shape.
Local masses in those correlators take similar values .
They are

$$
am(0^-)=0.174(38) ,  \ \ 0.169(61)  , \ \ 0.162(82) ,  \eqno[8].
$$

in SU(2), U$_1$ and the monopole part, respectively,  at $t=6 \sim 10$.

Although chiral limit ($ m \rightarrow 0 $) has not been examined,
existence of light pseudoscalar meson in those gauge field is confirmed.

On the other hand,
that in the regular part $\theta_\mu (n)^{Ph}$ behave differently as shown in
Fig.2 . In order to understand this behavior,
we present pseudoscalar correlator of free-staggered fermion in Fig.3 .
Saw tooth like behavior of the correlator in the regular gauge field is similar
to that of free staggered quark.  Therefore
in the regular gauge field, there seems no light pion and no evidence
for chiral symmetry breaking.

\section{Topological charge in the photon and the monopole part}

   Here we report an evidence  of strong correlation between instantons
and monopoles.  Relationship between instantons in nonabelian gauge
field and monopoles in its $U_1$ sector has not well clarified.
Nevertheless , intimate relation is expected due to several reasons.
Both are non-trivial topological objects. Instanton is known to couple
with chiral symmetry breaking.  Actually, we have shown
that monopole seems
relevant to the breaking in the previous work \cite{Miya}
and in this paper.

   Quantities examined here are charge $Q$, abolute integral of topological
charge density $I_Q$ and a normalized action $\tilde s$
defined by

$$
    Q={1 \over 32\pi^2} \sum \epsilon_{\mu \nu \lambda \sigma}
tr[ P_{\mu \nu}P_{\lambda \sigma}] , \eqno[9]
$$

$$
    I_Q={1 \over 32\pi^2} \sum |\epsilon_{\mu \nu \lambda \sigma}
tr[ P_{\mu \nu}P_{\lambda \sigma}]| , \eqno[10]
$$

$$
    \tilde s= {1 \over 8\pi^2 } \sum (1-{1 \over 2} tr[P_{\mu \nu}]) ,
\eqno[11]
$$

where $P_{\mu \nu}$ is a plaquette in $\mu \nu$ -plane.
We evaluate those quantities for the Ph-part and the Ds-part as well as
original SU(2) link variables with using cooling method.

   Since, $Q$ and $I_Q$ are those of SU(2) object, we reconstruct
SU(2) link variables $U_{\mu}^{Ph}(n)$ and $U_{\mu}^{Ds}(n)$
from $\theta^{Ph}_{\mu}(n)$ and $\theta^{Ds}_{\mu}(n)$ by multipling
the matter part $M_{\mu}(n)$
in the following way;

$$
U_{\mu}^{i}(n) = \left[\matrix{r_c(n)  & -c_{\mu}(n)
\cr c_{\mu}(n)^* &
r_c(n) }\right ] \times
\left[ \matrix{ e^{i\theta^{i}_{\mu}(n)} & 0 \cr  0 &
e^{-i\theta^{i}_{\mu}(n)} } \right ] , \eqno[11]
$$

where  $i$ takes either  Ph or Ds .

An sample of 40 configurations, we cool those three fields, $U_{\mu}(n)$ ,
$U_{\mu}^{Ph}(n)$,$U_{\mu}^{Ds}(n)$ and monitor $Q$ , $I_Q$ and $\tilde s$.
Fig.4a,4b and 4c show typical examples of cooling curves.
 Although there is
quantitative difference, $Q$ , $I_Q$ and $\tilde s$ of
SU(2) and Ds-part have nontrivial plateaus while those of Ph-part  die
very quickly.
The trivial feature in Ph-part is common in all 40 configurations.
Fig.5a and 5b show correlation between topological charges of SU(2)
and
Ph-part or Ds-part at 3 cooling.
Fig.s 6a and 6b are similar one but that at 100 cooling.
We see that a strong correlation between SU(2) and Ds-part while Ph-part
is trivial and no correlation.  Similar features are observed also in $I_Q$
and $\tilde s$. These suggest that Ds-part, i.e. monopole part ,
is responsible to topological charge and has close connection to instantons.

\section{Summary and conclusion}
Pseudoscalar correlator is examined in the maximal projected U$_1$ gauge field
and in its decomposed fields.
The correlators in SU(2), U$_1$ and $\theta_\mu (n)^{Ds}$ shows
similar exponential decay and existence of light pion mode.
On the other hand, in the regular part, the correlator behaves as a product
of free staggered fermion.  In addition to previous results
on chiral condensate at finite temperature \cite{Miya},
these results give support for an
idea that monopole condensation is also responsible for chiral symmetry
breaking.

   We find an evidence  of strong correlation between instantons
and monopoles.  The field dominated by monopoles carrys topological charge
similar to the original SU(2) field .  On the other hand ,
The field dominated by photons shows no topological charge.

Present results suggest possibility that
suitable effective theory of monopole condensation
can describe both chiral symmetry breaking of QCD as well as confinement.
\cite{Yotu}

\topskip 0cm
\section{Acknowledgements}
The authors acknowledge T.Suzuki, H.Toki, H.Suganuma, S.Hioki and S.Kitahara
for discussions and comments.

%\end{document}

\pagebreak
\epsfile{file=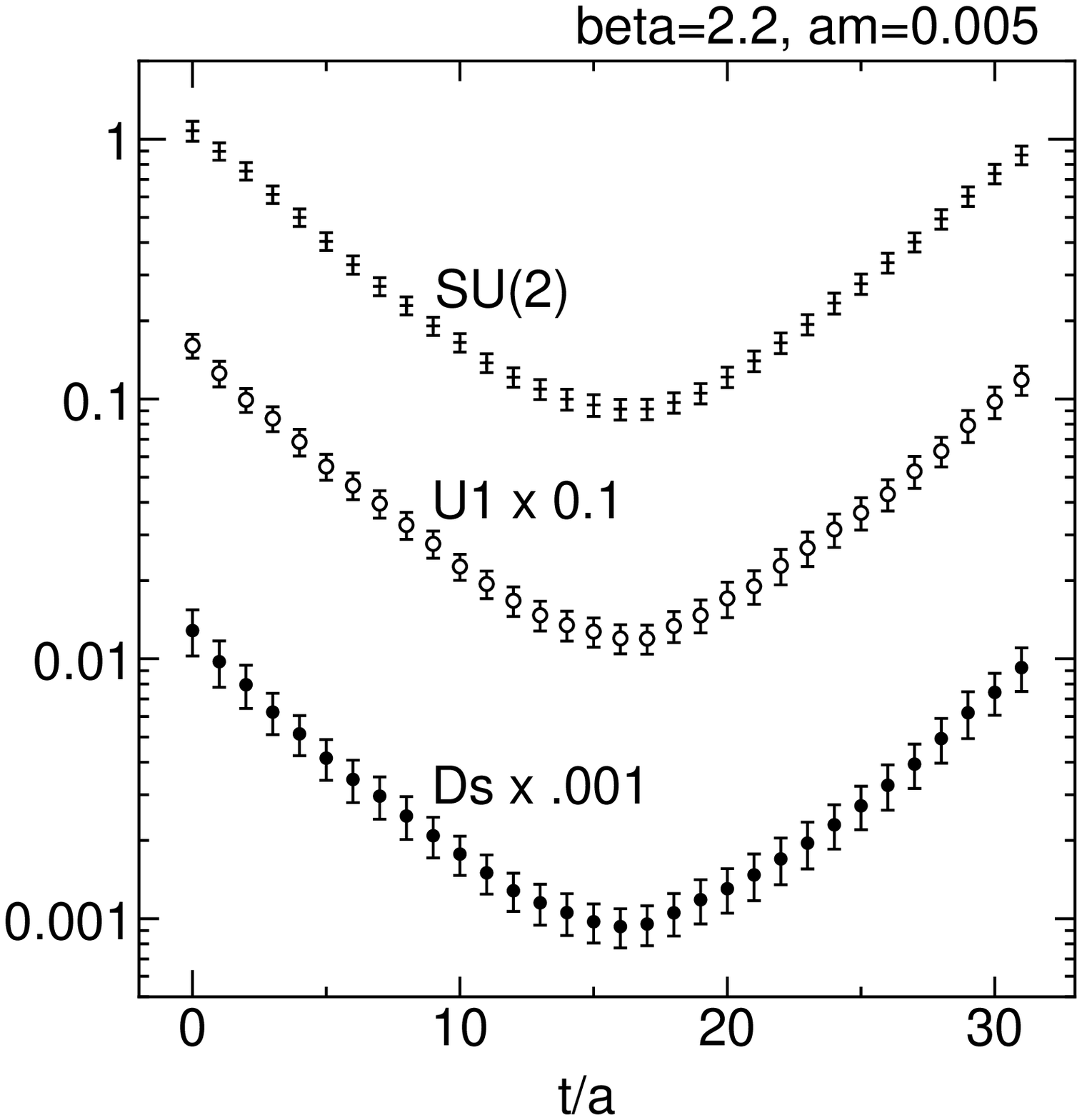,scale=0.8}

Fig.1 Pseudoscalar correlators in the SU(2), U$_1$ and monopole field.

\begin{itemize}
\item[]
Lattice is $16^3 \times 32$ at beta=2.2 and staggered quark mass
parameter is $ma=0.005$.
Crosses (SU(2)) ,
open circles ( U$_1$ ) , filled circles ( Monopole-part ) and
triangles ( Photon-part ) represent respective data.  \cite{Miya}
\end{itemize}

\epsfile{file=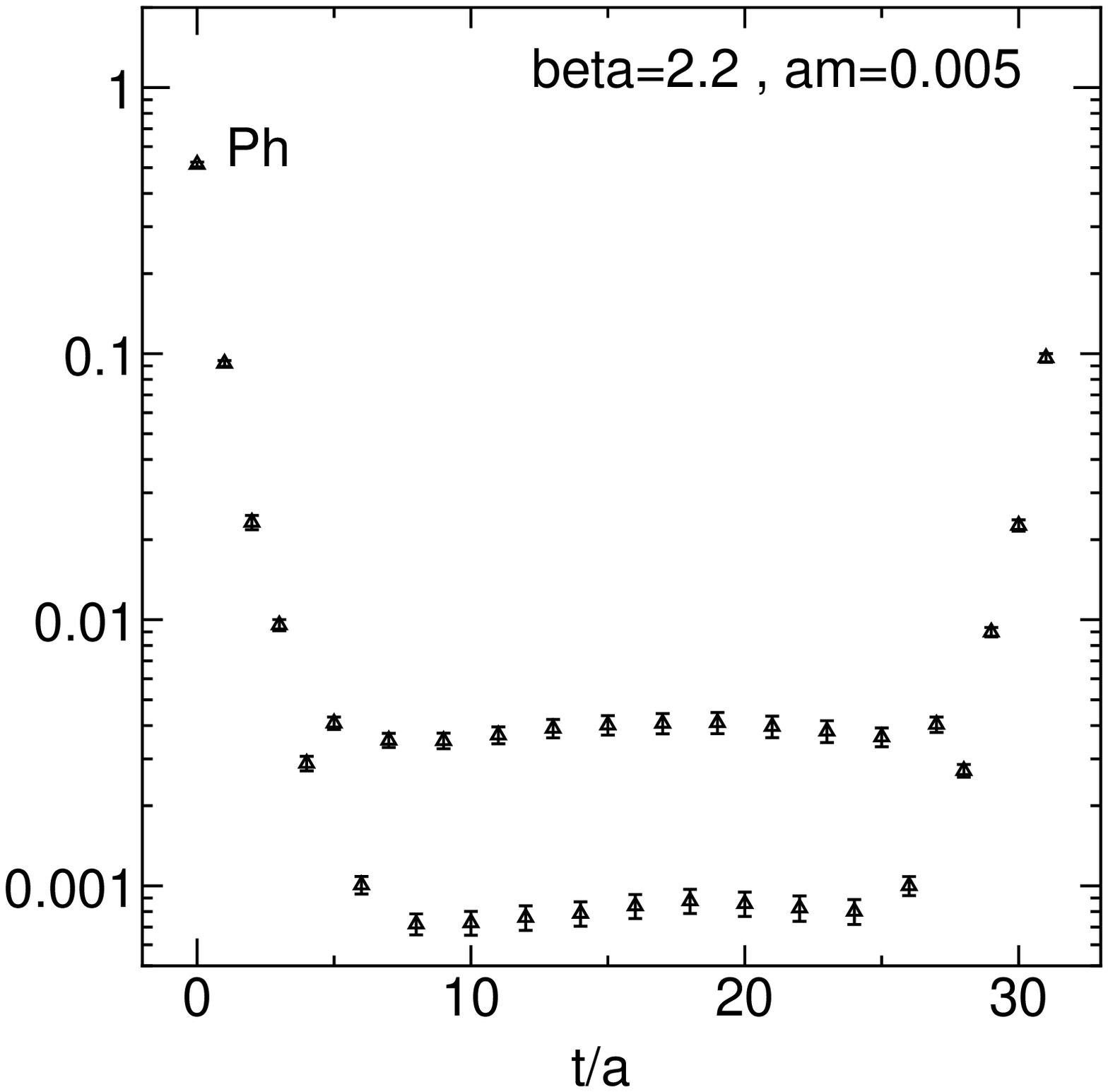,scale=0.8}

Fig.2 Pseudoscalar correlators in the photon field.

\begin{itemize}
\item[]
Lattice and staggered quark mass parameter are the same with Fig.1.
\end{itemize}

\epsfile{file=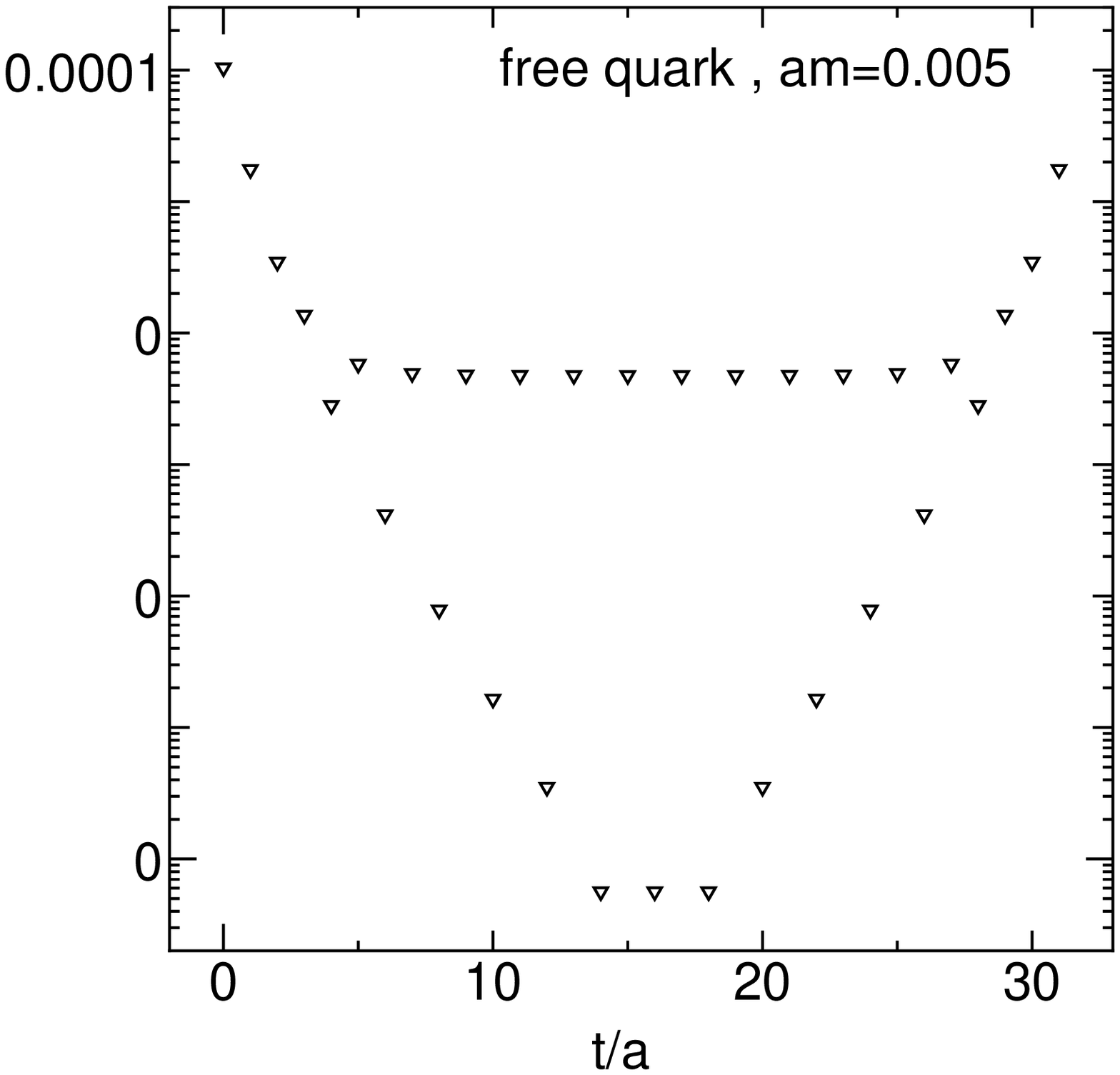,scale=0.8}

Fig.3 Pseudoscalar correlators by product of free staggered quark and
antiquark.

\begin{itemize}
\item[]
Lattice and staggered quark mass parameter are the same with Fig.1.
\end{itemize}

\epsfile{file=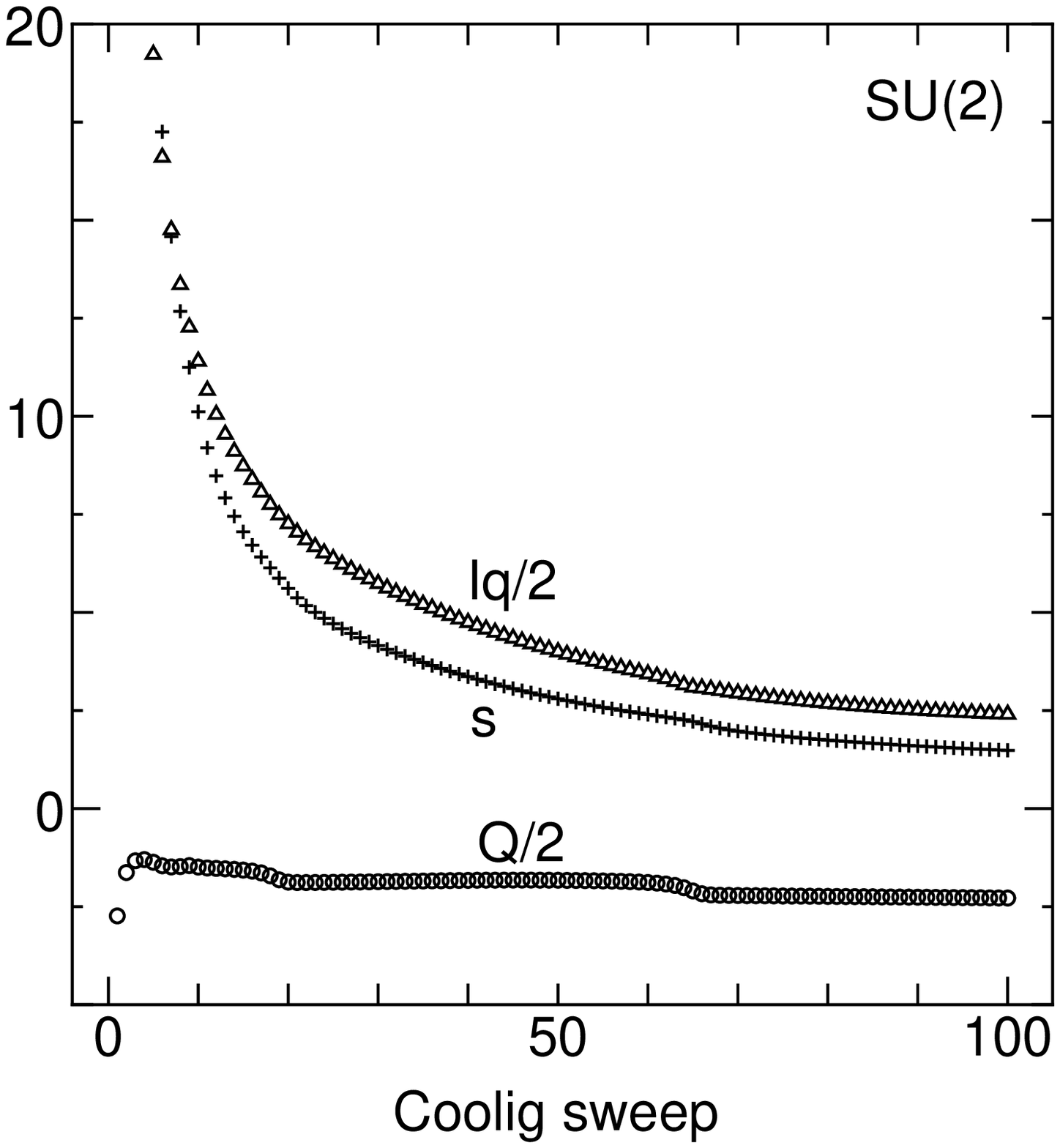,scale=0.8}

Fig.4a Typical example of cooling curves for a SU(2) gauge field.

\begin{itemize}
\item[]
Lattice is $16^4$ beta=2.4.
\end{itemize}

\epsfile{file=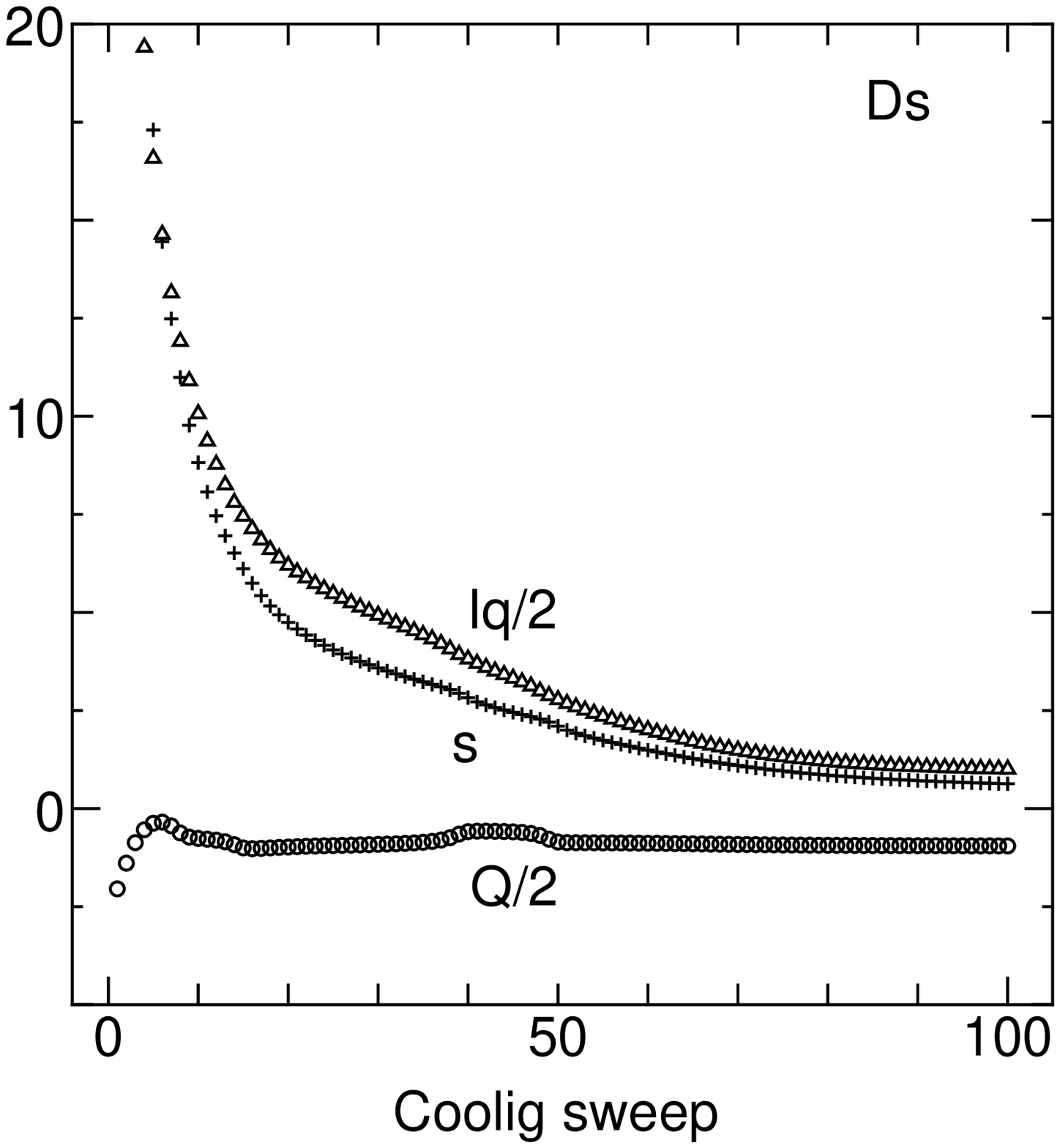,scale=0.8}

Fig.4b Cooling curves for a gauge field constructed from
Ds part. Original SU(2) configuration is the same one in Fig.4a.

\epsfile{file=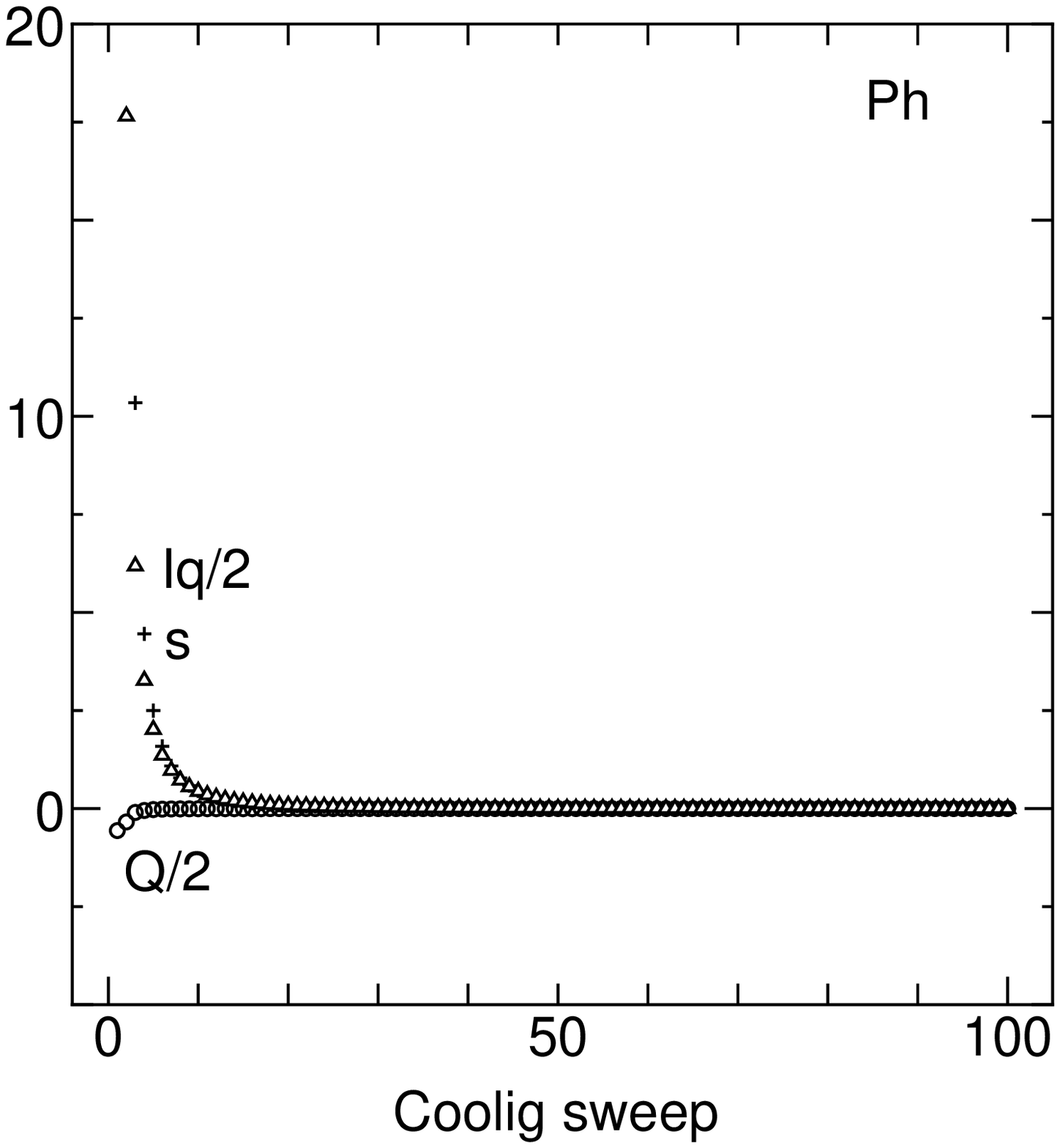,scale=0.8}

Fig.4c Cooling curves for a gauge field constructed form Ph part.
Original SU(2) configuration is the same one in Fig.4a.

\epsfile{file=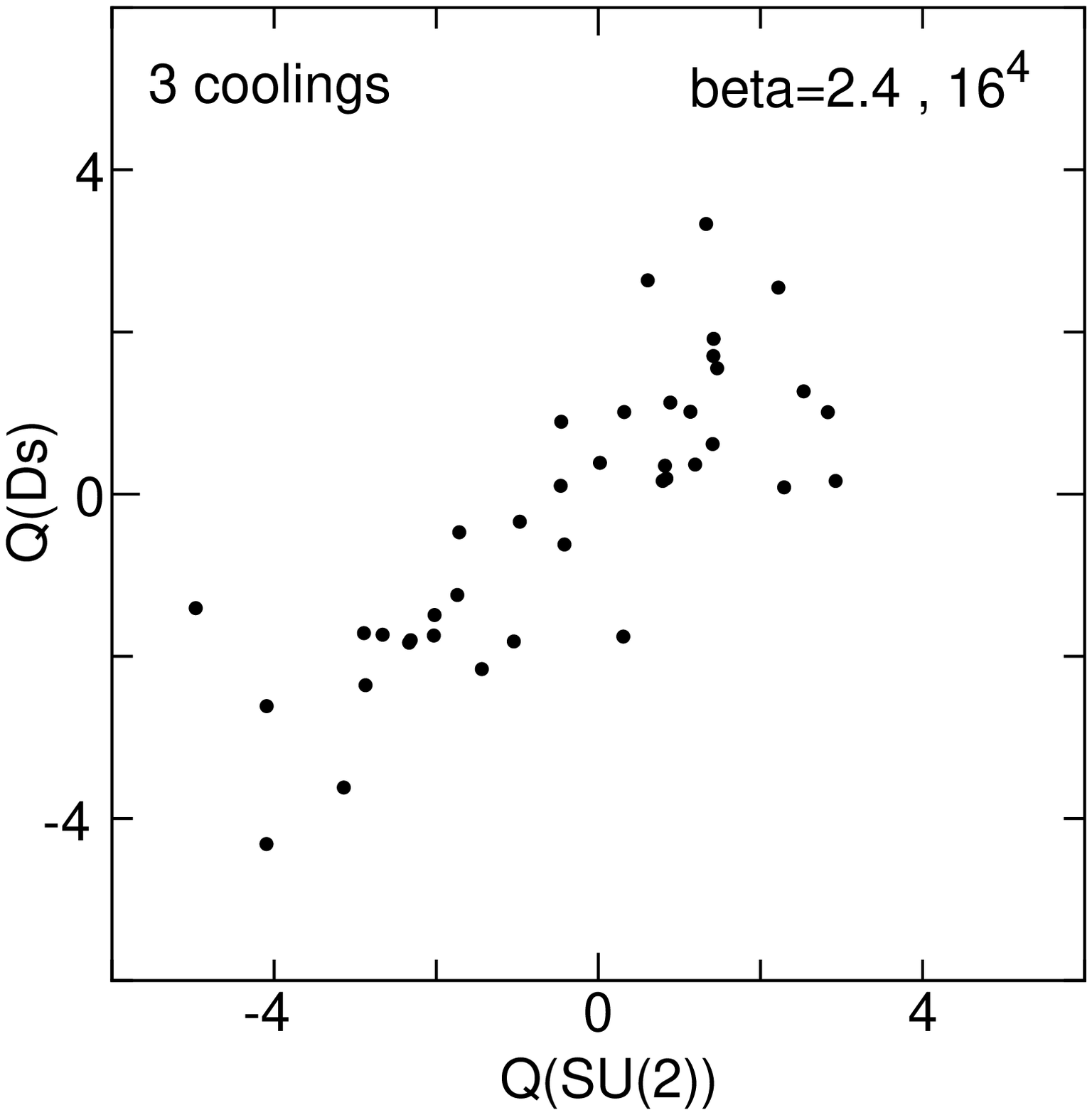,scale=0.8}

Fig.5a Correlation between $Q(SU(2))$ and $Q(Ds)$ at 3 cooling.
Number of data is 40. Lattice is $16^4$ at $\beta=2.4$.

\epsfile{file=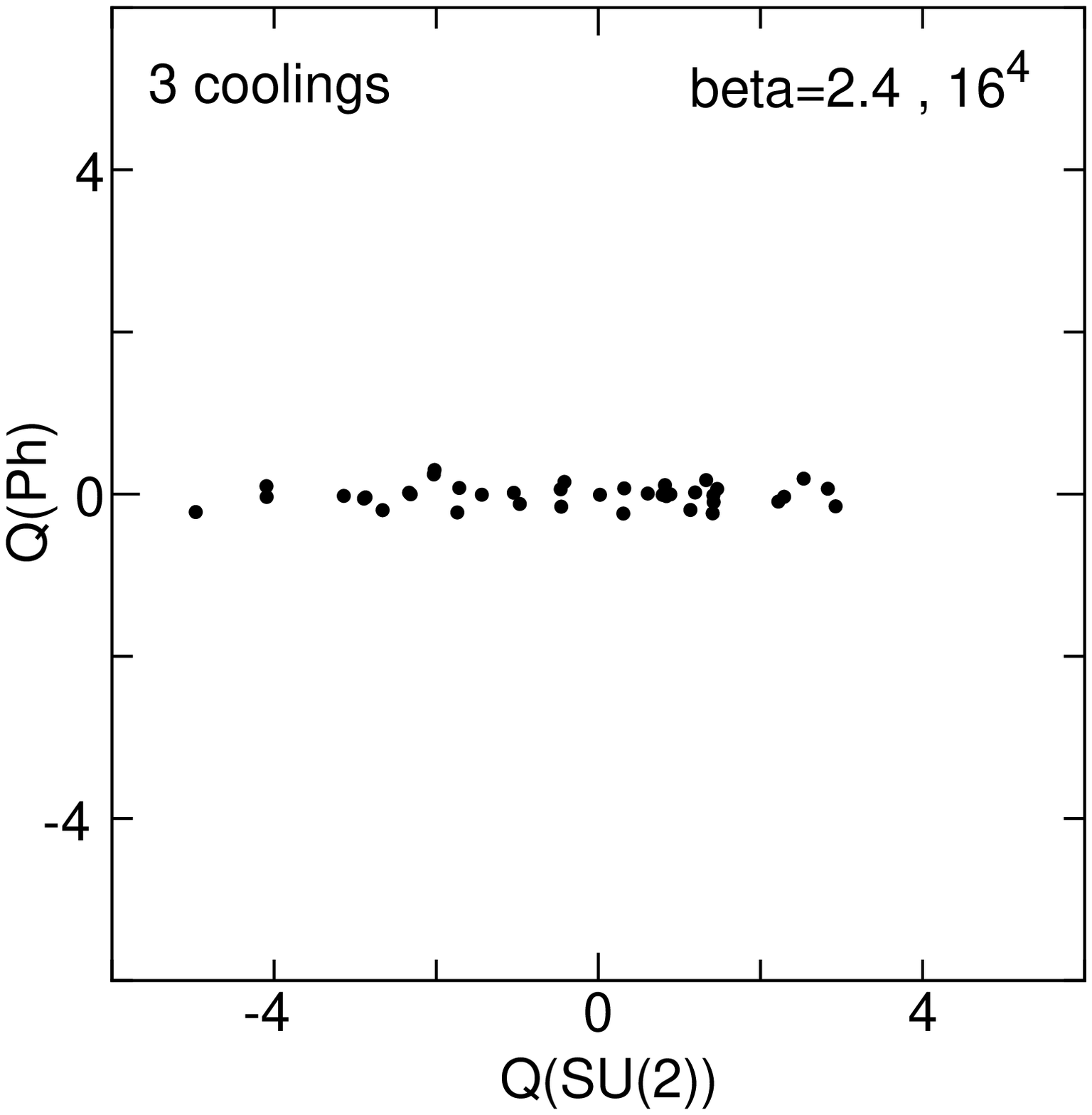,scale=0.8}

Fig.5b Correlation between $Q(SU(2))$ and $Q(Ph)$ at 3 cooling.

\epsfile{file=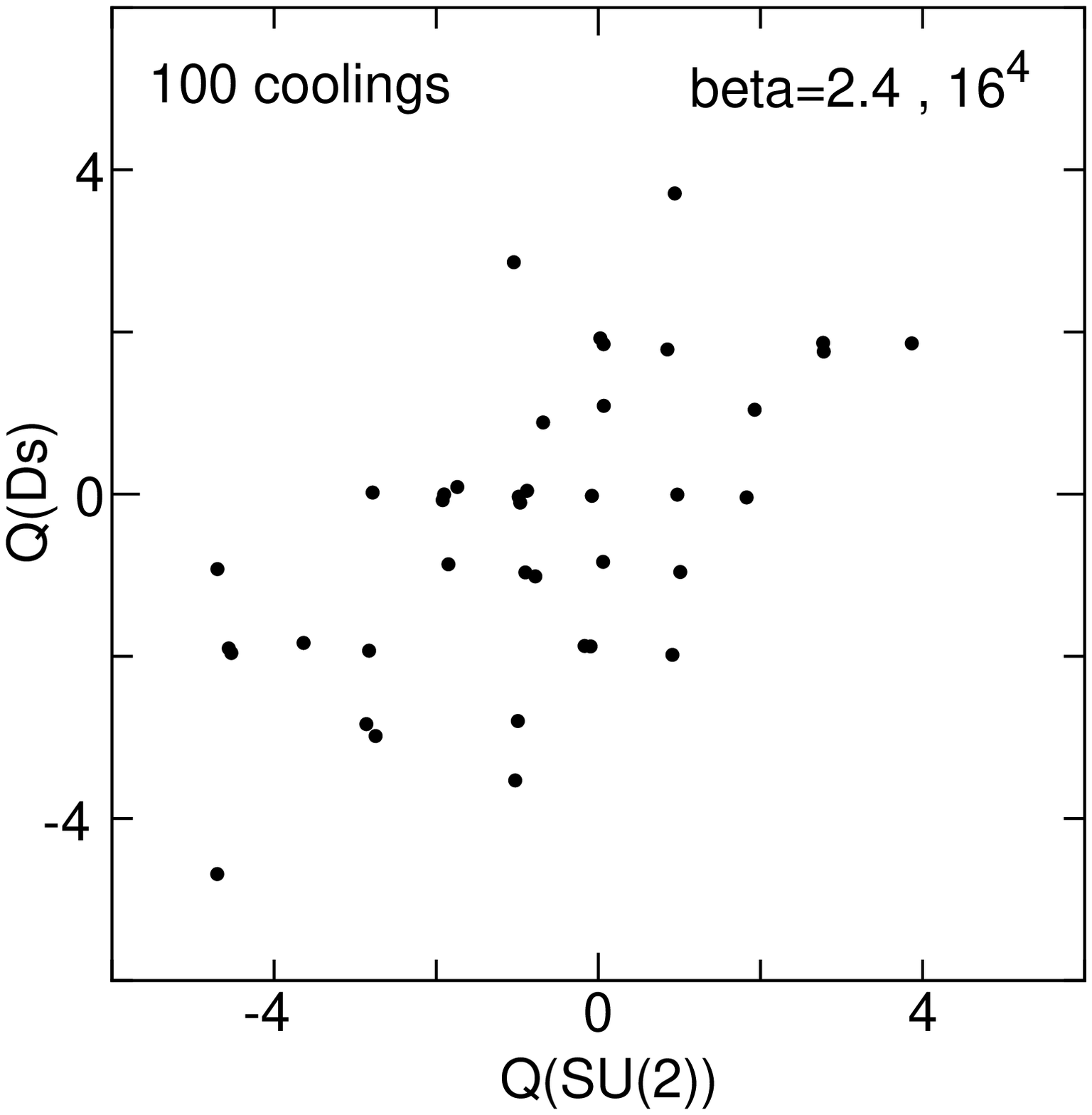,scale=0.8}

Fig.6a Correlation between $Q(SU(2))$ and $Q(Ds)$ at 100 cooling.

\epsfile{file=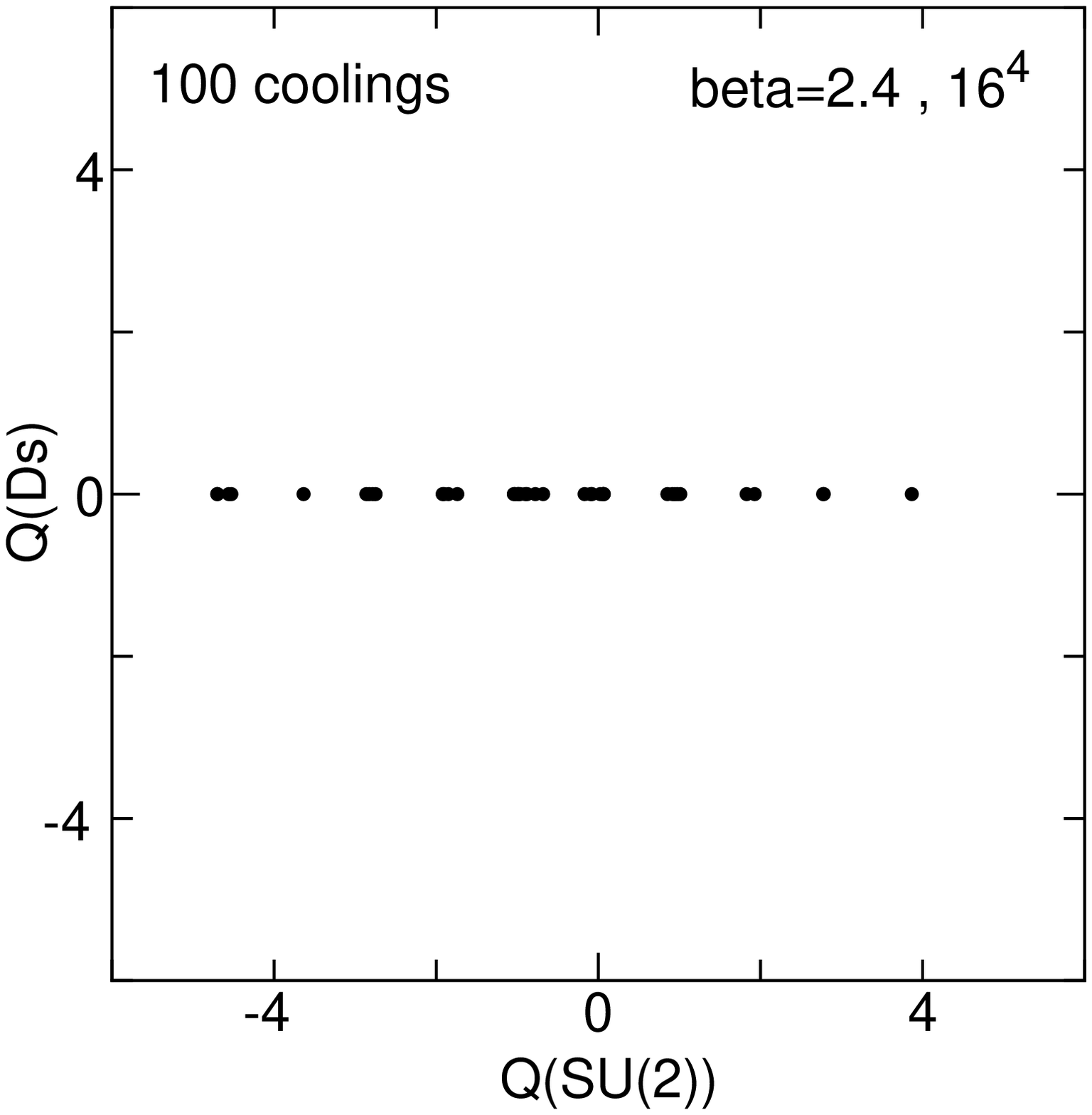,scale=0.8}

Fig.6b Correlation between $Q(SU(2))$ and $Q(Ph)$ at 100 cooling.


\section{References}
\begin{thebibliography}{9}
\bibitem{Miya} O.Miyamura, Nucl.Phys. B(Proc. Suppl.) 42 (1995) 538 ;
Phys.Letters B353 (1995) 91 .
\bibitem{Wolo} R.M.Woloshyn, TRIUMF preprint TRI-PP-94-80 (1994).
\bibitem{Kron} A.S.Kronfeld, M.L.Laursen and G.Schierholz. et al.,
Phys.Lett. {\bf 198B} (1987) 516.
\bibitem{Brands} F.Brandstaeter,U.J.Wiese and G.Schierholz ,
Phys.Lett. {\bf B272} (1991) 319.
\bibitem{Hioki} S.Hioki et al., Phys.Lett. {\bf B272} (1991) 326.
\bibitem{Ohno} S.Ohno et al., Nucl. Phys. {\bf B(Proc. Suppl.)30} (1993) 561.
\bibitem{Shiba} H.Shiba and T.Suzuki,Nucl. Phys. {\bf B(Proc. Suppl.) 34}
(1994) 182.
\bibitem{Giacomo} A. DiGiacomo, talk presented in the LATTCE94 at Bielfeld
Sept,(1994).
\bibitem{Casher} T. Banks and A. Casher, Nucl. Phys. {\bf B169} (1980) 103.
\bibitem{Baker} M.Baker, J.S.Ball and F.Zachariasen, Phys.Rev. {\bf D38} (1988)
1926;Erratum-ibid {\bf D47} (1993) 743.
\bibitem{Klein} G.Klein and Williams, Phys. Rev. {\bf D43} (1991) 3541.
\bibitem{Kamizawa} S.Kamizawa, Doctor Thesis, Kanazawa Univ. (1993).
\bibitem{Suga} H.Suganuma,S.Sasaki and H.Toki, Nucl. Phys. {\bf B435} (1995)
207 and references cited therein.
\bibitem{tHooft2} G. tHooft , Nucl. Phys. {\bf B190 (1981)} 455.
\bibitem{Matsu1} Y.Matsubara et al., talk presented in the LATTCE94 at Bielfeld
Sept,(1994).
\bibitem{Suzuki} T.Suzuki, preprint KANAZAWA 94-15 (hep-lat9408003).
\bibitem{DeG} T. DeGrand and D. Toussaint, Phys. Rev. {\bf D22} (1980) 2478.
\bibitem{Forc} Ph.de Forcrand , talk presented in the LATTCE94 at Bielfeld
Sept,(1994).
\bibitem{Yotu} I.Yotsuyanagi and T.Suzuki, Phys. Rev. {\bf D42} (1990) 4257.
\end{thebibliography}
\end{document}